\documentclass[aps,prl,twocolumn,showpacs,superscriptaddress,groupedaddress]{revtex4-1}
\usepackage{color}
\usepackage{graphicx}
\usepackage{amsmath}			
\usepackage{rotating}			
\usepackage{setspace}
\usepackage{here}
\usepackage[german,english]{babel}
\usepackage{bbm}			

\usepackage{natbib}

\usepackage{comment}

\excludecomment{comt}

\begin{document}
%
%
%
\def\bdm{\begin{displaymath}}
\def\edm{\end{displaymath}}
\def\nn{\nonumber}
\def\bc{\begin{center}}
\def\ec{\end{center}}
\def\be{\begin{equation}}
\def\ee{\end{equation}}
\def\tcb{\textcolor{blue}}
\def\tcbl{\textcolor{black}}
\def\tcg{\textcolor{green}}
\def\tcr{\textcolor{red}}
\def\tcgr{\textcolor{grey}}
\def\va{{\bf a}} \def\vA{{\bf A}} \def\vb{{\bf b}} \def\vB{{\bf B}} \def\vb{{\bf b}} \def\vc{{\bf c}}
\def\vC{{\bf C}} \def\vd{{\bf d}} \def\hvd{\hat\vd} \def\vD{{\bf D}} \def\ve{{\bf e}} \def\hve{\hat\ve}
\def\vE{{\bf E}} \def\vf{{\bf f}} \def\vF{{\bf F}} \def\vg{{\bf g}} \def\vG{{\bf G}} \def\vh{{\bf h}}
\def\vH{{\bf H}} \def\vi{{\bf i}} \def\vI{{\bf I}} \def\vj{{\bf j}} \def\vJ{{\bf J}} \def\vk{{\bf k}}
\def\hvk{\hat\vk} \def\vK{{\bf K}} \def\vl{{\bf l}} \def\vL{{\bf L}} \def\vLambda{{\bf\Lambda}}
\def\vm{{\bf m}} \def\vM{{\bf M}} \def\vn{{\bf n}} \def\hvn{\hat\vn} \def\vN{{\bf N}} \def\vone{{\bf 1}}
\def\vp{{\bf p}} \def\hvp{\hat\vp} \def\vP{{\bf P}} \def\vq{{\bf q}} \def\vQ{{\bf Q}} \def\vr{{\bf r}}
\def\vR{{\bf R}} \def\vs{{\bf s}} \def\vS{{\bf S}} \def\vt{{\bf t}} \def\vT{{\bf T}} \def\vu{{\bf u}}
\def\vU{{\bf U}} \def\vv{{\bf v}} \def\vV{{\bf V}} \def\vw{{\bf w}} \def\vW{{\bf W}} \def\vx{{\bf x}}
\def\vX{{\bf X}} \def\vy{{\bf y}} \def\vY{{\bf Y}} \def\vz{{\bf z}} \def\v0{{\bf 0}} \def\hvz{\hat\vz}
\def\vZ{{\bf Z}} \def\vtau{{\bf \tau}} \def\e{{\rm e}} \def\kB{k_{\rm B}} \def\kF{k_{\rm F}} \def\EF{E_{\rm F}}
\def\ra{{\rm a}} \def\rb{{\rm b}} \def\rc{{\rm c}} \def\rh{{\rm h}}
\def\NF{N_{\rm F}} \def\pF{p_{\rm F}} \def\Tc{T_{\rm c}} \def\vvF{v_{\rm F}} \def\vna{{\bf\nabla}}
\def\vPi{{\bf\Pi}} \def\Tc{T_{\rm c}} \def\muB{\mu_{\rm B}} \def\Nf{N_{\rm F}}
\def\lela{\left\langle} \def\rira{\right\rangle} \def\Tc{T_{\rm c}}
%
%
\def\vxi{{\mbox{\boldmath$\xi$}}}
\def\vlambda{{\mbox{\boldmath$\lambda$}}}
\def\vDelta{{\mbox{\boldmath$\Delta$}}}
\def\vna{{\mbox{\boldmath$\nabla$}}}
\def\vtau{{\mbox{\boldmath$\tau$}}}
\def\vDelta{{\mbox{\boldmath$\Delta$}}}
\def\vsigma{{\mbox{\boldmath$\sigma$}}}
\def\vepsilon{{\mbox{\boldmath$\varepsilon$}}}
\def\vPsi{{\mbox{\boldmath$\Psi$}}}
\def\vvarphi{{\mbox{\boldmath$\varphi$}}}
\def\vchi{{\mbox{\boldmath$\chi$}}}
\def\valpha{{\mbox{\boldmath$\alpha$}}}
\def\vbeta{{\mbox{\boldmath$\beta$}}}
\def\vgamma{{\mbox{\boldmath$\gamma$}}}
\def\vrho{{\mbox{\boldmath$\rho$}}}
\def\vmu{{\mbox{\boldmath$\mu$}}}
\def\vpsi{{\mbox{\boldmath$\psi$}}}
\def\vGamma{{\mbox{\boldmath$\Gamma$}}}
%

\def\<{\left\langle}	\def\>{\right\rangle}
\def\d{\delta}
\def\l{\lambda}
\def\d{\delta}
\newcommand{\+}{\dagger}
\title{\bf Leggett modes and the Anderson--Higgs mechanism\\
in superconductors without inversion symmetry}
%
\author{Nikolaj Bittner$^1$, Dietrich Einzel$^2$, Ludwig Klam$^1$ and Dirk Manske$^1$\\
$^1$Max--Planck--Institut f\"ur Festk\"orperforschung, D--70569 Stuttgart, Germany\\
$^2$Walther--Meissner--Institut f\"ur Tieftemperaturforschung, D--85748 Garching, Germany
}
\def\va{{\bf a}} \def\vA{{\bf A}} \def\vb{{\bf b}} \def\vB{{\bf B}} \def\vb{{\bf b}} \def\vc{{\bf c}}
\def\vC{{\bf C}} \def\vd{{\bf d}} \def\hvd{\hat\vd} \def\vD{{\bf D}} \def\ve{{\bf e}} \def\hve{\hat\ve}
\def\vE{{\bf E}} \def\vf{{\bf f}} \def\vF{{\bf F}} \def\vg{{\bf g}} \def\vG{{\bf G}} \def\vh{{\bf h}}
\def\vH{{\bf H}} \def\vi{{\bf i}} \def\vI{{\bf I}} \def\vj{{\bf j}} \def\vJ{{\bf J}} \def\vk{{\bf k}}
\def\hvk{\hat\vk} \def\vK{{\bf K}} \def\vl{{\bf l}} \def\vL{{\bf L}} \def\vLambda{{\bf\Lambda}}
\def\vm{{\bf m}} \def\vM{{\bf M}} \def\vn{{\bf n}} \def\hvn{\hat\vn} \def\vN{{\bf N}} \def\vone{{\bf 1}}
\def\vp{{\bf p}} \def\hvp{\hat\vp} \def\vP{{\bf P}} \def\vq{{\bf q}} \def\vQ{{\bf Q}} \def\vr{{\bf r}}
\def\vR{{\bf R}} \def\vs{{\bf s}} \def\vS{{\bf S}} \def\vt{{\bf t}} \def\vT{{\bf T}} \def\vu{{\bf u}}
\def\vU{{\bf U}} \def\vv{{\bf v}} \def\vV{{\bf V}} \def\vw{{\bf w}} \def\vW{{\bf W}} \def\vx{{\bf x}}
\def\vX{{\bf X}} \def\vy{{\bf y}} \def\vY{{\bf Y}} \def\vz{{\bf z}} \def\v0{{\bf 0}} \def\hvz{\hat\vz}
\def\vZ{{\bf Z}} \def\vtau{{\bf \tau}} \def\e{{\rm e}} \def\kB{k_{\rm B}} \def\kF{k_{\rm F}} \def\EF{E_{\rm F}}
\def\NF{N_{\rm F}} \def\pF{p_{\rm F}} \def\Tc{T_{\rm c}} \def\vvF{v_{\rm F}} \def\vna{{\bf\nabla}}
\def\vPi{{\bf\Pi}} \def\Tc{T_{\rm c}}
\def\lela{\left\langle} \def\rira{\right\rangle} \def\Tc{T_{\rm c}}
%
%
\def\vxi{{\mbox{\boldmath$\xi$}}} \def\vlambda{{\mbox{\boldmath$\lambda$}}} \def\vDelta{{\mbox{\boldmath$\Delta$}}}
\def\vna{{\mbox{\boldmath$\nabla$}}} \def\vtau{{\mbox{\boldmath$\tau$}}} \def\vDelta{{\mbox{\boldmath$\Delta$}}}
\def\vsigma{{\mbox{\boldmath$\sigma$}}} \def\vkappa{{\mbox{\boldmath$\kappa$}}}
\begin{abstract}
{We develop a microscopic and gauge--invariant theory for collective modes 
resulting from the phase of the superconducting order parameter in 
non--centrosymmetric superconductors. Considering various crystal symmetries we derive the corresponding 
gauge mode $\omega_{\rm G}(\vq)$ and find, in particular, new Leggett modes 
$\omega_{\rm L}(\vq)$ with characteristic properties that are unique to 
non--centrosymmetric superconductors.
We calculate their mass and dispersion that reflect the underlying
spin--orbit coupling and thus the balance between triplet and singlet superconductivity occurring
simultaneously. 
Finally, we demonstrate the role of the Anderson--Higgs mechanism: while the 
long--range Coulomb interaction shifts $\omega_{\rm G}(\vq)$ to 
the condensate plasma mode $\omega_{\rm P}(\vq)$, it leaves the mass $\Lambda_0$ of the new Leggett mode
unaffected and only slightly modifies its dispersion.
}
\end{abstract}


 \maketitle
{\it Introduction.} 
Owing to the Pauli exclusion principle in single--band superconductors spin--singlet 
(even parity) and triplet (odd parity) pairing correlations never occur simultaneously. 
Important examples are spin--triplet odd--parity pairing correlations in superfluid 
$^3$He~\cite{LEGGETT1975,VWBOOK1990}, triplet superconductivity in Sr$_2$RuO$_4$~\cite{MANDM2003}, 
as well as unconventional singlet pairing correlations in heavy Fermion systems~\cite{PFLEIDERER2009} and
cuprates~\cite{NANDP2003}. 
A necessary prerequisite for a clear singlet--triplet distinction is, however, the existence of 
an inversion center. The discovery of the bulk superconductors CePt$_3$Si (tetragonal~\cite{TETNCS}) 
and Li$_2$Pd$_x$Pt$_{3-x}$B (cubic~\cite{CUBNCS:01, *CUBNCS:02}), without inversion symmetry, 
to give only two examples, has therefore initiated extensive theoretical and experimental studies. 
The Rashba--type spin--orbit coupling caused by the absence 
of an inversion center implies (i) the lifting of the band degeneracy associated with a splitting 
into a two--band structure and (ii) the superposition of both singlet and triplet contributions 
to the superconducting gap~\cite{ASOC:01, *ASOC:02, *ASOC:03, NCS2012}.

The breaking of a continuous symmetry in superconductors 
is associated with the occurrence of a gauge mode which is necessary 
to restore the charge conservation. Furthermore, in analogy to the Josephson effect, 
Leggett predicted the appearance of a new collective excitation 
in $s$--wave two--band superconductors, which corresponds to an out--of--phase oscillation 
mode of the phase difference of the coupled condensates~\cite{LEGGETT1966}. So far, the Leggett mode 
has been only observed in MgB$_2$~\cite{Blumberg:2007}, but several predictions for other 
$s$--wave superconductors 
have been made~\cite{ZEHETMAYER2013, Ota:2011, LinHu:2012}. In non--centrosymmetric superconductors (NCS), 
however, where a complex mixing of singlet and triplet 
superconductivity occurs, it is not {\it a priori} clear whether a Leggett mode exists~\cite{NCS2012}. 

In this letter we use a microscopic theory to demonstrate the existence of Leggett modes in NCS. 
For this purpose, we calculate all order parameter collective modes associated with 
the condensate phase dynamics. For the first time we provide analytic expressions
and numerical calculations for the gauge mode $\omega_{\rm G}$, 
the mass and the dispersion of Leggett's collective mode 
$\omega_{\rm L}(\vq)$
, as well as for the frequency $\omega_{\rm P}$ of the condensate plasma mode. 
The interesting interplay of these collective modes is studied in connection 
with the electromagnetic response 
of the pair condensate, with special emphasis on the participation of the collective modes in the
Anderson--Higgs mechanism \cite{ANDERSON1963,HIGGS1964}. We emphasize the calculation of the mass $\Lambda_0$
of various Leggett modes that depend strongly on the singlet--to--triplet ratio and may be observable 
by Raman or Brillouin scattering experiments.

{\it Model description of NCS.} The Hamiltonian for noninteracting electrons 
in a non--centrosymmetric crystal reads
\begin{equation}
\hat{\cal H}=\sum_{\vk\sigma\sigma^\prime}\hat c^\dagger_{\vk\sigma}
\left[ \xi_\vk \delta_{\sigma\sigma^\prime}
+\vgamma_\vk\cdot\vtau_{\sigma\sigma^\prime}\right] \hat c_{\vk\sigma^\prime}\ ,
\end{equation}
where $\xi_\vk$ represents the bare band dispersion, $\sigma,\sigma^\prime={\uparrow,\downarrow}$ label 
the spin state and $\vtau$ are the Pauli matrices. The second term describes an 
antisymmetric spin--orbit coupling (ASOC) through the vector $\vgamma_\vk$. In NCS two important classes of ASOCs are realized 
which reflect the underlying point group $\mathcal{G}$ of the crystal. 
We shall particularly be interested in the tetragonal point group $C_{4v}$ 
(relevant for CePt$_3$Si) and the cubic point group $O(432)$ 
(applicable to the system Li$_2$Pd$_x$Pt$_{3-x}$B). For $\mathcal{G}=C_{4v}$ the ASOC reads
\begin{equation}
\vgamma_\vk=\mathrm{\gamma}_\bot ( \hat\vk\times\hat\ve_z ) + 
\mathrm{\gamma}_\Vert \hat k_x\hat k_y\hat k_z ( \hat k^2_x - \hat k^2_y ) \hat\ve_z\ .
\label{eq:C4v}
\end{equation}
In the purely 2D case ($\mathrm{\gamma}_\Vert=0$) one recovers the 
Rashba interaction. For the cubic point group $\mathcal{G}=O(432)$ $\vgamma_\vk$ reads
$\vgamma_\vk = \mathrm{\gamma}_1 \hat \vk -\mathrm{\gamma}_3 
[ \hat k_x (\hat k_y^2 + \hat k_z^2)\hat\ve_x
+ \hat k_y (\hat k_z^2 + \hat k_x^2)\hat\ve_y
+ \hat k_z (\hat k_x^2 + \hat k_y^2)\hat\ve_z ]\ .$

What are the consequences of the ASOC? First, diagonalizing the Hamiltonian, 
one finds the energy eigenvalues 
$\xi_{\vk\mu}=\xi_\vk+\mu||\vgamma_\vk||$
with $\mu=\pm1$ which correspond to a lifting of the band degeneracy between 
the two spin states at a given momentum $\hbar\vk$. This band splitting is responsible for the 
two--band structure characteristic of NCS metals. Second, the presence of an ASOC invalidates the 
classification of the superconducting order parameter with respect to spin singlet (even parity)
and spin triplet (odd parity). Thus, in general, a linear combination of the gap on both bands is 
possible. Sigrist and co-workers have shown that most likely $\vgamma_\vk$
orientates parallel to the $\vd$--vector of the triplet part~\cite{ASOC:02, *ASOC:03}. 
Thus, we can simply write the gap function on the two bands in terms 
of a singlet ($\Delta_s$) and a triplet ($\Delta_{tr}$) amplitude:
\begin{equation}
\Delta_{\vk\mu}=\Delta_s(T)+\mu\Delta_{tr}(T)f_\vk\ ,
\label{eq:GapNCS}
\end{equation}
with
$
\ f_\vk=||\vgamma_\vk||/[\lela||\vgamma_{\vk^\prime}||^2\rira_{\rm FS}]^{1/2}\geq0 
$, where $\<\dots\>_{\rm FS}$ denotes the Fermi surface (FS) average~\cite{FS}. 
Thus, in short, while for all superconductors having an inversion center {\it either} singlet 
{\it or} triplet pairing is realized, in NCS singlet {\it and} triplet pairing occurs
simultaneously. Simply speaking, the resulting ASOC may drive e.g. $s$-- plus $p$--wave pairing on 
one band while $s$-- minus $p$--wave is established on the other,
leading to new collective modes.

{\it Nonequilibrium Kinetic Theory for NCS.} 
In order to calculate the dynamical properties of NCS
we consider the response to 
a scalar electromagnetic potential $\phi(\vq,\omega)$. 
In addition there contributes a charge fluctuation term, 
which accounts for the action of the 3D long--range Coulomb
interaction $V_\vq=4\pi e^2/\vq^2$ within the RPA, i.e. $\chi=\chi^{(0)}[1-V_\vq\chi^{(0)}]^{-1}$,
where $\chi$ is a generalized response function. Then,
the response to the 
perturbation $\d\zeta\equiv e\phi(\vq,\omega)+V_\vq\d n(\vq,\omega)$ with $\d n$ being 
the total density response of the
system, is described by a generalized momentum distribution function 
$\underline{n}_{\vp\vp^\prime}^{\mu}$
which is a $2\times2$--matrix in Nambu--space using again the band basis of Eq.~\eqref{eq:GapNCS}
with $\mu=\pm1$. 
\begin{figure}[!t]
 \centering
  \includegraphics[width=0.76\columnwidth,clip]{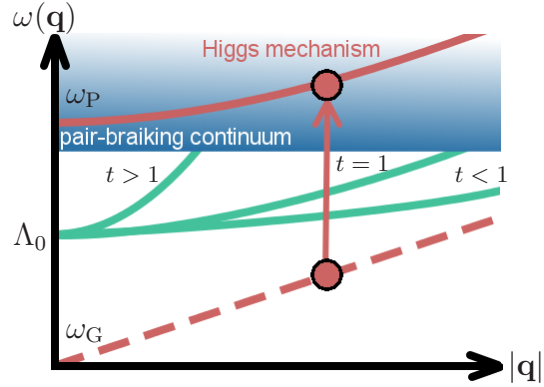}
  \put(0,3){\large $|\vq|$}
  \put(-187,135){\large $\omega(\vq)$}
  \put(-187,52){\large $\Lambda_0$}
  \put(-20,76){\small $t<1$}
  \put(-65,80){\small $t=1$}
  \put(-152,76){\small $t>1$}
  \put(-167,17){\large $\omega_{\rm G}$}
  \put(-167,106){\large $\omega_{\rm P}$}
\caption{(color online) Illustration of various calculated collective modes (T=0) common to all NCS.
The Anderson--Higgs mechanism shifts the gauge mode $\omega_{\rm G}$ (dashed line)
to the plasma mode $\omega_{\rm P}$ usually lying in the 
pair--breaking continuum. The new Leggett modes (solid green lines) unique to NCS
are only slightly changed by this process (not visible) and the mass $\Lambda_0$ remains unchanged.
Importantly, in some cases $\Lambda_0\to0$ is possible, see discussion of Fig.~\ref{fig:LM009}; thus,
the Leggett modes might be easy observable. Note that the slope of the Leggett modes depend on the
ratio $t=\Delta_{tr}/\Delta_s$ as discussed in connection with Fig.~\ref{fig:LDispC4v}. 
}
\label{fig:DispGMLM}
\end{figure}
At the same time the perturbation $\delta\zeta$ induces fluctuations 
$\d g_{\vk\mu}^{(-)}\equiv\frac{1}{2}[\d g_{\vk\mu}\frac{\Delta_{\vk\mu}^*}{|\Delta_{\vk\mu}|}
-\frac{\Delta_{\vk\mu}}{|\Delta_{\vk\mu}|}\d g_{-\vk\mu}^*]$ of the pairing amplitude $g_{\vk\mu}$,
as well as the important phase fluctuations of the superconducting order parameter
$\d \Delta_{\vk\mu}^{(-)}\equiv\frac{1}{2}[\d\Delta_{\vk\mu}\frac{\Delta_{\vk\mu}^*}{|\Delta_{\vk\mu}|}
-\frac{\Delta_{\vk\mu}}{|\Delta_{\vk\mu}|}\d\Delta_{-\vk\mu}^*]$, which we will later use to 
determine all collective modes.
The Fourier transformation of $\underline{n}_{\vp\vp^\prime}^\mu$ describes the evolution of the 
system in space and time after perturbation $\d\zeta$. However, it is convenient to stay
in $(\vq, \omega)$--space and solve the von Neumann equation~\cite{NB:01}
\begin{equation}
  \hbar\omega\,\underline{n}_{\vp\vp^\prime}^\mu+\sum_{\vp^{\prime\prime}}
  [\underline{n}_{\vp\vp^{\prime\prime}}^\mu,\,\underline{\xi}_{\vp^{\prime\prime}\vp^{\prime}}^{\mu}]=0
\label{eq:MKE}
\end{equation}
in the clean limit, where $\vp=\hbar\left(\vk+\vq/2\right)$, $\vp^\prime=\hbar\left(\vk-\vq/2\right)$ 
and the $2\times2$ energy matrix $\underline{\xi}_{\vp^{\prime\prime}\vp^\prime}^\mu$ have been 
introduced. The simplest way to solve Eq.~\eqref{eq:MKE} is to make the following ansatz:
\begin{equation}
 \begin{array}{llllclc}
  \underline{n}_{\vp\vp^\prime}^{\mu}&\equiv&\underline{n}_{\vk\mu}(\vq, \omega)&
  =&\underline{n}_{\vk\mu}^0\d_{\vq,0}&+&\d\underline{n}_{\vk\mu}(\vq, \omega)\\
  \underline{\xi}_{\vp\vp^\prime}^\mu&\equiv&\underline{\xi}_{\vk\mu}(\vq, \omega)&
  =&\underline{\xi}_{\vk\mu}^0\d_{\vq,0}&+&\d\underline{\xi}_{\vk\mu}(\vq, \omega)  
 \end{array}
\label{eq:Lin}
\end{equation} with the nonequilibrium quantities
{\small \begin{equation*}
\label{eq:DefDistFkt}
\delta\underline{n}_{\vk\mu}=\left(
 \begin{array}{cc}
  \d n_{\vk\mu} &\mu\d g_{\vk\mu}\\
  \mu\d g_{\vk\mu}^*&-\d n_{-\vk\mu}
 \end{array}
 \right),\delta\underline{\xi}_{\vk\mu}=\left(
 \begin{array}{cc}
  \d \zeta &\mu \d\Delta_{\vk\mu}\\
  \mu\d\Delta_{\vk\mu}^*&-\d\zeta
 \end{array}
 \right)\ .
\end{equation*}} After some lengthy, but straightforward calculations [supplement 
material, Eqs.~(A.7)-(A.11)] we obtain from 
the off--diagonal components of Eq.~\eqref{eq:MKE} the relation between fluctuations of the 
pairing amplitude and fluctuations of the superconducting order parameter:
%
%
%
%
{\small \begin{equation}
\label{eq:floatedEquation}
\begin{split}
2\Delta_{\vk\mu}[\delta g_{\vk\mu}^{(-)}+&\theta_{\vk\mu}\delta\Delta_{\vk\mu}^{(-)}]
=\\
&\omega\lambda_{\vk\mu}\delta\zeta-[\omega^2-(\vq\cdot\vv_{\vk\mu})^2]\lambda_{\vk\mu}
\frac{\delta\Delta_{\vk\mu}^{(-)}}{2\Delta_{\vk\mu}}\ .
\end{split}
\end{equation}}
Here, we have identified the condensate response function
{\small
\begin{equation}
\lambda_{\vk\mu}={4\Delta_{\vk\mu}^2}\frac{\theta_{\vk\mu}[\omega^2-(\vq\cdot\vv_{\vk\mu})^2]
+\Phi_{\vk\mu}(\vq\cdot\vv_{\vk\mu})^2}
{(\vq\cdot\vv_{\vk\mu})^2[\omega^2-4\xi_{\vk\mu}^2]-\omega^2[\omega^2-4E_{\vk\mu}^2]}\label{eq:DefTsuneto}
\end{equation}
}with $\vv_{\vk\mu}=\partial\xi_{\vk\mu}/\partial \hbar\vk$,
$\theta_{\vk\mu}=\tanh(E_{\vk\mu}/{2\kB T})/2E_{\vk\mu}$, 
$E_{\vk\mu}=[\xi_{\vk\mu}^2+\Delta_{\vk\mu}^2]^{1/2}$
and  
$\Phi_{\vk\mu}=-\partial n_{\vk\mu}/\partial\xi_{\vk\mu}$ with momentum distribution 
function $n_{\vk\mu}$. An important property of the condensate response is the sum rule, 
which generates the condensate density
$
\sum_{\vp\mu}\lambda_{\vp\mu}=
N_0\sum_{\mu}\lela\lambda_{\hat\vp\mu}\rira_{\rm FS}\equiv N_0\lambda\ ,
$
with $N_0=\NF/2$ being the DoS for one spin projection.
As we will show in supplement material [Eqs.~(A.16)-(A.18)] 
the total particle density $\delta n$ 
obeys the conservation law
$
\omega\delta n-\vq\cdot\vj=0
$
only, if all phase fluctuation modes of the order parameter in Eq.~\eqref{eq:floatedEquation}
are properly accounted for.

Finally, we find from the diagonal components of Eq.~\eqref{eq:MKE} the density response of NCS:
{\small
\begin{equation}
\delta n_{\vk\mu}=\left(\frac{\left(\vq\cdot\vv_{\vk\mu}\right)^2\varphi_{\vk\mu}}
{\omega^2-\left(\vq\cdot\vv_{\vk\mu}\right)^2}-\lambda_{\vk\mu}\right)\delta\zeta
+\omega\lambda_{\vk\mu}\frac{\delta\Delta_{\vk\mu}^{(-)}}{2\Delta_{\vk\mu}}
\label{eq:DiagEq}
\end{equation}
}with $\varphi_{\vk\mu}=\Phi_{\vk\mu}-\lambda_{\vk\mu}$ being the quasiparticle response.
Since we are only interested in the response of the superconducting condensate $\d n_{\rm s}$, 
we may ignore quasiparticle contributions $\propto\varphi_{\vk\mu}$ in Eq.~\eqref{eq:DiagEq}.
Then, the density response function simplifies to
$
\ \delta n_{\vk\mu}~=~-\lambda_{\vk\mu}\delta\zeta
+\omega\lambda_{\vk\mu}\delta\Delta_{\vk\mu}^{(-)}/2\Delta_{\vk\mu}
$.
Hence, the condensate density response
$\delta n_{\rm s}=\sum_{\vk\mu}\delta n_{\vk\mu}$
is exclusively determined by $\lambda_{\vk\mu}$. In other words, we find that the frequency-- 
and wave--vector dependence of $\d n_{\rm s}(\vq,\omega)$ contains all information
on the relevant order parameter collective modes in NCS. Finally, combining 
Eqs.~\eqref{eq:floatedEquation} with both the superconducting gap equation 
$\Delta_{\vk \mu} = \sum_{\vp\nu}\Gamma_{\vk\vp}^{\mu\nu}g_{\vp\nu}$ and its variation
\begin{equation}
\delta\Delta_{\vk\mu}^{(-)}=\sum_{\vp\nu}\Gamma_{\vk\vp}^{\mu\nu}\delta g_{\vp\nu}^{(-)}
\label{eq:VarGapEq}
\end{equation}
(with $\Gamma_{\vk\vp}^{\mu\nu}$ being the pairing interaction~\footnote{Since the pairing interaction 
$\Gamma_{\vk\vp}^{\mu\nu}$ occurs in both gap equation and in Eq.~\eqref{eq:VarGapEq} 
describing Cooper--pair phase fluctuations it is possible to eliminate $\Gamma_{\vk\vp}^{\mu\nu}$ while 
inserting both Eqs. in Eq.~\eqref{eq:floatedEquation}.}) 
leads to the main result of our analysis ($\hat\vq=\vq/|\vq|$):
\begin{widetext}
\begin{eqnarray}
\label{eq:Response}
\d n_{\rm s}(\vq,\omega)=N_0\l
\frac{\omega_{\rm G}^2(\vq)[\omega^2-\omega_{\rm L}^{\prime2}(\vq)]}{\omega^4-
\left[\omega_{\rm P}^2(\hat\vq)+\omega_{\rm G}^2(\vq)
+\omega_{\rm L}^2(\vq)\right]\omega^2+\left[\omega_{\rm P}^2(\hat\vq)+\omega_{\rm G}^2(\vq)\right]
\omega_{\rm L}^{\prime 2}(\vq)} e\phi(\vq,\omega)\ .
\end{eqnarray}
\end{widetext}

{\it New collective modes.} From the denominator of Eq.~\eqref{eq:Response} we can 
draw important conclusions which are summarized in Fig.~\ref{fig:DispGMLM}. 
In analogy to neutral systems we first consider $\omega_{\rm P}(\hat{\vq})\to0$
and find two poles 
{\small\tcbl{\begin{equation}
\begin{array}{cll}
\label{eq:PolUnrenorm}
  \omega_1^2=&\omega_{\rm G}^2(\vq)+{\cal O}\left(\frac{\omega_{\rm G}^4(\vq)}
  {\omega_{\rm L}^2(\vq)}\right)&\quad \text{gauge mode}\\
  \omega_2^2=&\omega_{\rm L}^2(\vq)+{\cal O}\left(\frac{\omega_{\rm G}^4(\vq)}
  {\omega_{\rm L}^2(\vq)}\right)&\quad \text{Leggett mode}
\end{array}
\end{equation}}}with $\omega_{\rm G}(\vq)$
being the characteristic {\it gauge} mode of NCS with
$
\omega_{\rm G}^2(\vq)=\sum_{\mu}\lela\lambda_{\hat\vp\mu}(\vq\cdot\vv_{\vp\mu})^2\rira_{\rm FS}/\lambda\ .
$
Furthermore, we discover the Anderson--Higgs mechanism for the gauge mode in NCS 
shifting it to the plasma frequency, i.e.  
$\omega_{\rm P}^2(\vq)=\omega_{\rm P}^2(\hat\vq)+\omega_{\rm G}^2(\vq)$. Thus,
after Coulomb renormalization, we find:
\tcbl{{\small\begin{equation}
\begin{array}{cll}
\label{eq:PolRenorm}
  \omega_1^2=&\omega_{\rm P}^2(\hat\vq)+\omega_{\rm G}^2(\vq)
+{\cal O}\left(\frac{\omega_{\rm L}^2(\vq)}
  {\omega_{\rm P}^2(\hat\vq)}\right)&\text{plasma mode}\\
  \omega_2^2=&\omega_{\rm L}^{\prime 2}(\vq)+{\cal O}\left(\frac{\omega_{\rm L}^2(\vq)}
  {\omega_{\rm P}^2(\hat\vq)}\right)&\text{Leggett mode}
\end{array}
\end{equation}}with $\omega_{\rm P}(\hat\vq)$ being the characteristic }
{\it condensate plasma frequency} of NCS with
$
\omega_{\rm P}^2(\hat\vq)=4\pi n e^2\sum_\mu 3\lela\lambda_{\hat\vp\mu}(\hat\vq\cdot\hat\vp)^2\rira_{\rm FS}/m
$. 
It is important to note that the
full condensate density response $\d n_{\rm s}$ as described by Eq.~\eqref{eq:Response} 
is also manifested in the condensate dielectric function
$\epsilon\equiv1-V_\vq\d n_{\rm s}^{(0)}/e\phi$, with $\d n_{\rm s}^{(0)}\equiv\d n_{\rm s}
(\omega_{\rm P}(\hat\vq)\to0)$.
All in all, our new results for the gauge mode and plasma frequency generalizes 
the known solutions for ordinary two--band superconductors which can be 
obtained in the limit $f_\vk\equiv 1$~\cite{Sharapovetal:2002}.

\begin{figure}[!t]
  \centering
  \includegraphics[width=0.75\columnwidth,clip]{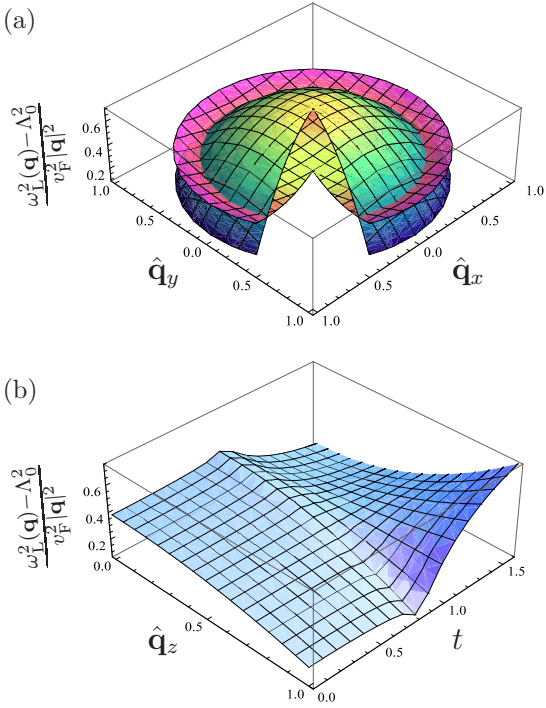}
  \put(-205,255){(a)}
  \put(-205,115){(b)}  
  \put(-150,20){\large $\hat{\vq}_z$}
  \put(-35,20){\large $t$}
  \put(-150,160){\large $\hat{\vq}_y$}
  \put(-35,160){\large $\hat{\vq}_x$}
  \put(-185,50){\large \begin{rotate}{95}$\frac{\omega_{\rm L}^2(\vq)-\Lambda_0^2}{v_{\rm F}^2|\vq|^2}$
  \end{rotate}}
  \put(-185,187){\large \begin{rotate}{95}$\frac{\omega_{\rm L}^2(\vq)-\Lambda_0^2}{v_{\rm F}^2|\vq|^2}$
  \end{rotate}}
\caption{(color online) Slope of the dispersion 
of the Leggett mode for NCS systems with $C_{4v}$ point group symmetry
as a function of the unit vectors $\hat{\vq}_x, \hat{\vq}_y, \hat{\vq}_z$.
(a) Comparison of the slope for $t=0.5$ (upward parabola) and $t=1.5$ (downward parabola),
(b) slope along the $\hat{\vq}_z$--direction for various $t=\Delta_{tr}/\Delta_s$.
}
\label{fig:LDispC4v}
\end{figure}

The second pole in Eq.~\eqref{eq:Response} leads with  $\omega_{\rm P}(\hat{\vq})\to0$
to Eq.~\eqref{eq:PolUnrenorm} 
 determining the new {\it Leggett's collective modes} $\omega_{\rm L}(\vq)$ in NCS corresponding to 
oscillations in the relative phase of the superconducting condensates. 
The exact analytical result 
for $\omega_{\rm L}^2(\vq)$ is too lengthy to be shown here and thus can be found in 
the supplement material [see Eqs. (B3)-(B5)]. Instead,
we illustrate its dispersion (for different $t=\Delta_{tr}/\Delta_s$) in Fig.~\ref{fig:DispGMLM} 
and calculate its slope (as an example for $C_{4v}$) in Fig.~\ref{fig:LDispC4v}. As expected, 
we find for all point groups considered the dispersion 
$\left(\omega_{\rm L}^2(\vq)-\Lambda_0^2\right)\propto |\vq|^2$. 
The slope, however, depends on the ratio $t=\Delta_{tr}/\Delta_s$. Thus, in Fig.~\ref{fig:LDispC4v}(a)
we show the slope of the Leggett mode exemplarily for the tetragonal point group $C_{4v}$ 
[see Eq.~\eqref{eq:C4v}] along the $\hat{\vq}_x$-- and $\hat{\vq}_y$--direction. The calculated upward 
parabola corresponds
to $t=0.5$ while the downward parabola corresponds to $t=1.5$, respectively. 
For $t=1$ one finds a constant slope of $1/3$ 
(independent of $\hat{\vq}_x$ and $\hat{\vq}_y$, not shown). 
The three resulting dispersions
are illustrated schematically in Fig.~\ref{fig:DispGMLM}. In Fig.~\ref{fig:LDispC4v}(b) 
we show the slope along the $\hat{\vq}_z$--direction for various $t$ which reveals 
a non--monotonic behavior for fixed $\hat{\vq}_z$.
In contrast, for the cubic point group $O(432)$ we find in all directions 
$\omega_{\rm L}^2(\vq)-\Lambda_0^2=\frac{1}{3} \vv_{\rm F}^2|\vq|^2$ {\it independent} of $t$ (not shown), 
since the underlying ASOC is isotropic to leading order. 
This would correspond to the curve with $t=1$ in Fig.~\ref{fig:DispGMLM}.
%

From $\omega_2^2(\vq)$ in Eq.~\eqref{eq:PolUnrenorm}, we find 
the mass $\Lambda_0$ of the Leggett mode
\begin{equation}
  \Lambda_0^2\equiv\omega_{\rm L}^2(\vq=0)=4\gamma_{\rm ncs}\Delta_s\Delta_{tr}
  \frac{\lambda}{\lambda_0\lambda_2-\lambda_1^2}\ , 
\label{eq:LeggettMass}
\end{equation}
where the definitions
\begin{equation}
\lambda_n=
\Delta_s\Delta_{tr}\sum_{\mu=\pm1}\lela\lambda_{\hat\vp\mu}(\mu f_\vp)^n/\Delta_{\vp\mu}^2\rira_{\rm FS}
\label{eq:DefLambda}
\end{equation}
have been used. Here, $\gamma_{ncs}$ represents the coupling strength of the Leggett mode, 
which we will calculate below.
\begin{figure}[!t]
\centering
\includegraphics[width=0.82\columnwidth,clip]{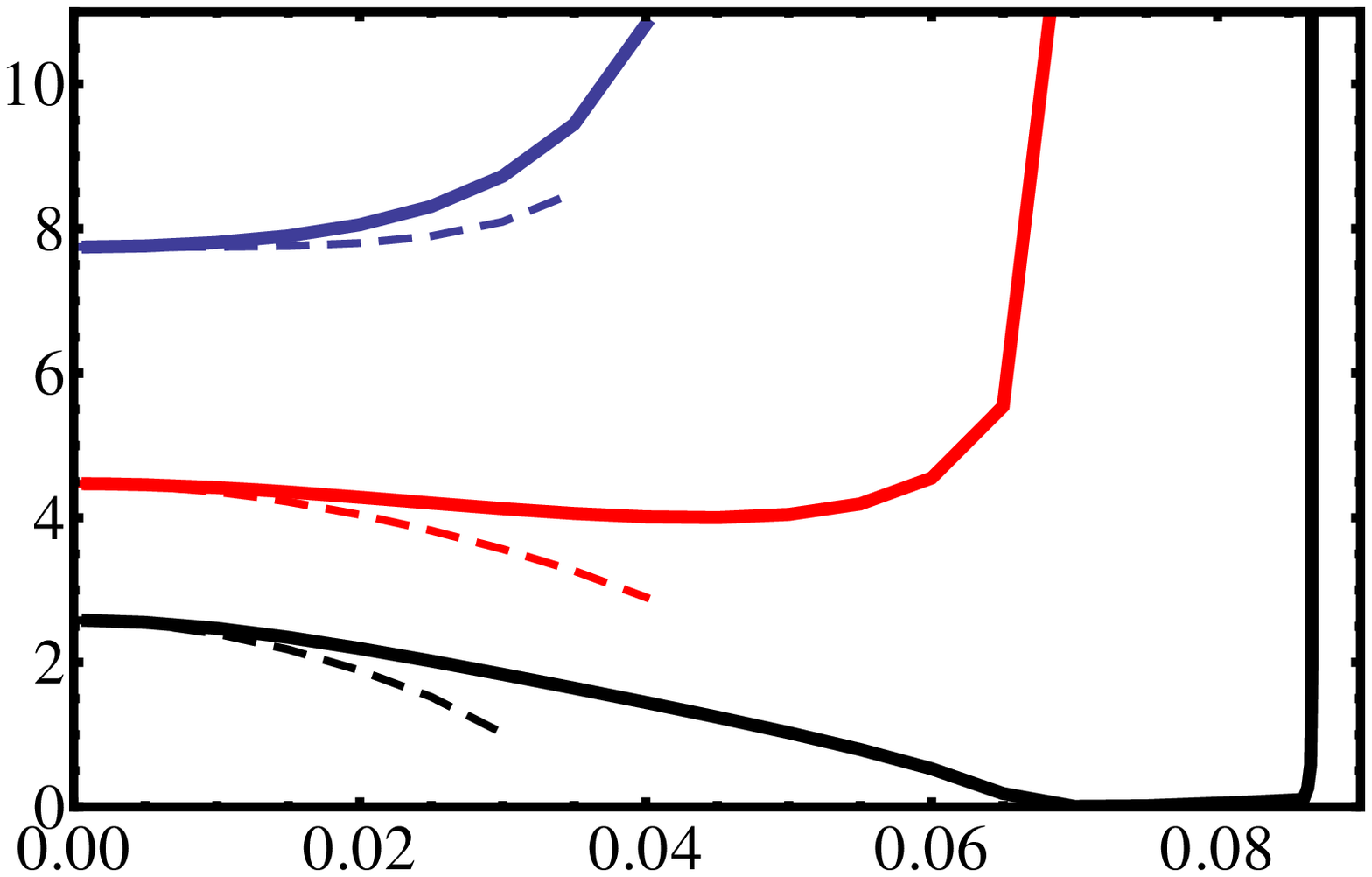}
  \put(-105,-12){\(\l_m\)}
  \put(-195,50){\begin{rotate}{90}\(\Lambda_0/\Delta_s\)\end{rotate}}  
  \put(-163,104){\(\l_{tr}=\frac{1}{4}\l_s\)}
  \put(-105,62){\(\l_{tr}=\frac{1}{2}\l_s\)}
  \put(-53,19){\(\l_{tr}=\frac{3}{4}\l_s\)}  
\caption{(color online) Normalized mass of the Leggett mode for NCS systems with $C_{4v}$ point group
symmetry for fixed $\l_s=0.1$ as a function of the mixing term and various $\l_{tr}$:
$\l_{tr}=0.025$ (upper solid line), $\l_{tr}=0.05$ (middle solid line) and 
$\l_{tr}=0.075$ (lower solid line). 
The dashed lines correspond to Eq.~\eqref{eq:LeggettMassAnalyt} which is an analytical solution
in the limit of small $t$.
}
\label{fig:LM009}
\end{figure}
In order to determine $\Lambda_0$ we need the {\it exact}
solution of the coupled self--consistency equations of the superconducting gap functions 
[see also Eq.~\eqref{eq:GapNCS}]: 
$\Delta_{\vk \mu} = \sum_{\vp\nu=\pm1}\Gamma_{\vk\vp}^{\mu\nu}g_{\vp\nu}$
with $g_{\vp\nu}=-\theta_{\vp \nu}\Delta_{\vp \nu}$ being the pairing amplitude and 
$\theta_{\vp\nu}$ has been defined together with Eq.~\eqref{eq:DefTsuneto}. We
choose the generalized two--gap weak--coupling pairing interaction of Ref.~\cite{SANDM2008}
$\Gamma_{\vk\vp}^{\mu\nu} = {\bf -}\left\{\Gamma_s + \Gamma_{tr} \mu\nu f_\vk f_\vp
+ \Gamma_m \left(\mu f_\vk + \nu f_\vp\right)\right\}\\
\Theta(\epsilon_0-|\xi_{\vk\mu}|) \Theta(\epsilon_0-|\xi_{\vp\nu}|)$ and obtain 
\begin{eqnarray}
\label{eq:EnGapEquilibMatrixForm}
\left\{-\vlambda^{-1}+
\left(\begin{array}{cc}
\Xi_0 & \Xi_1 \\
\Xi_1 & \Xi_2 \\
\end{array}\right)\right\}\cdot\left(\begin{array}{c}
\Delta_s  \\
\Delta_{tr}  \\
\end{array}\right)&=&\left(\begin{array}{c}
0 \\
0  \\
\end{array}\right) 
\end{eqnarray}
with $\lambda_\alpha~=~N(0)\Gamma_\alpha$, $\alpha~=~s,tr,m$,
\begin{eqnarray}
\label{eq:EnGapEquilibMatrixFormDef}
\vlambda =\left(\begin{array}{cc}
\lambda_s & \lambda_m \\
\lambda_m & \lambda_{tr} \\
\end{array}\right) \ ; \ \vlambda^{-1}=\frac{1}{|\vlambda|}\left(\begin{array}{cc}
\lambda_{tr} & -\lambda_m \\
-\lambda_m & \lambda_s \\
\end{array}\right)\ \ \ 
\end{eqnarray}
and $\Xi_n=\sum_{\mu}\lela\theta_{\hat\vp\mu}\left(\mu f_{\vp}\right)^n\rira_{\rm FS}$. 
Note that one obtains the ordinary two--band case if $\Xi_1\to0$~\cite{BANDE2013}. 
Equations~\eqref{eq:EnGapEquilibMatrixForm}--\eqref{eq:EnGapEquilibMatrixFormDef} have the advantage 
that the exact relation
\begin{equation}
  \gamma_{\rm ncs}=\lambda_m/|\vlambda|+\Xi_1 
\end{equation}
holds and thus determines the coupling constant in Eq.~\eqref{eq:LeggettMass}. Thus, for given
$\l_s, \l_{tr}, \l_m$ a numerical exact solution of Eq.~\eqref{eq:EnGapEquilibMatrixForm}
is always possible: the resulting exact gap function $\Delta_{\vk\mu}$ needs to be inserted in 
Eqs.~\eqref{eq:LeggettMass} and~\eqref{eq:DefLambda} to determine $\Lambda_0$~\cite{NB:03}.

In Fig.~\ref{fig:LM009} we show results for the Leggett mass $\Lambda_0$ for fixed $\l_s=0.1$ as a
function of $\l_m$. While for a small triplet contribution (upper solid line) $\Lambda_0$
increases monotonically, we find a non--monotonic behavior of the mass for increasing $\l_{tr}$ 
(middle solid line). Finally, if $\l_s\approx\l_{tr}$ we obtain the important case that
$\Lambda_0$ can become zero (lower solid line). Physically, this corresponds 
to a partly vanishing gap on one of the 
Fermi surfaces [see Eq.~\eqref{eq:GapNCS} and Ref.~\cite{KEM:2009}]. 
Also displayed in 
Fig.~\ref{fig:LM009} is the analytical solution in the limit of small $t$ 
(dashed lines) 
\begin{equation}
 \Lambda_0^2=2\Delta_s^2\frac{\l_s-\l_{tr}}{|\vlambda|}
 \left[1-\left(3\<f_\vk^4\>_{\rm FS}-1\right) t^2\right]\ .
 \label{eq:LeggettMassAnalyt}
\end{equation}
This might help
experimentalists to estimate in which materials the new Leggett modes are most easiest observable.

Finally, we return to the Anderson--Higgs mechanism. What is its role for the new Leggett modes? 
First, we conclude that the Leggett mass $\Lambda_0$ is unchanged, since the r.h.s. of 
Eq.~\eqref{eq:LeggettMass} does not depend on $\omega_{\rm G}$. Physically, this corresponds to 
the fact that the Meissner effect in the presence of a new Leggett mode is unchanged. 
Second, we find that the dispersion of the Leggett mode is only slightly changed. To see this, 
one needs to consider the difference in $\omega_2^2$ between Eqs.~\eqref{eq:PolUnrenorm} 
and~\eqref{eq:PolRenorm}. 
Since $\omega_{\rm G}(\vq\to0)\to0$ and $\omega_{\rm P}\gg\omega_{\rm L}$, the higher order corrections 
nearly vanish. The resulting $(\omega_{\rm L}^2 - \omega_{\rm L}^{\prime 2})$ is also very small 
[see supplement material Eq.~(C.4)]. Thus, we conclude that the dispersion of the Leggett mode and the results shown in 
Fig.~\ref{fig:LDispC4v} are nearly unchanged due to the Anderson--Higgs mechanism~\footnote{
A similar result has been found for the case
of ordinary two--band superconductors \cite{LEGGETT1966,Sharapovetal:2002,BITTNER2012}.}.
%

%
%
In conclusion, using a gauge--invariant theory of superconducting phase fluctuations in NCS 
we have demonstrated the existence of Leggett modes and calculated their characteristic mass 
and dispersion for various crystal symmetries. Both properties reflect the underlying spin--orbit coupling 
and depend strongly on the singlet--to--triplet ratio. Furthermore, we have calculated the corresponding 
gauge modes and clarified the role of the Anderson--Higgs mechanism for collective modes in NCS. 

{\it Acknoledgment.} The authors are grateful to R.~Gross and P.~Hirschfeld for helpful discussions.
\bibliographystyle{apsrev4-1}
\bibliography{BKEMprl.bib}

\begin{thebibliography}{31}
\bibitem [{FS()}]{FS}%
  \BibitemOpen
  \href@noop {} {}\bibinfo {note} {The average of the Fermi surface is defined
  as $\<z(\vk^\prime)\>_{\rm FS}=\int d\phi\int d\theta z(\vk^\prime)
  \sin\theta$ for a given function $z(\vk^\prime)$.}\BibitemShut {Stop}%
\end{thebibliography}

{\onecolumngrid
\begin{center}
\large{\bf Leggett modes and the Anderson--Higgs mechanism\\
in superconductors without inversion symmetry\\
{\bf Supplement material}
}\\
\ \newline
\normalsize{Nikolaj Bittner$^1$, Dietrich Einzel$^2$, Ludwig Klam$^1$ and Dirk Manske$^1$\\
$^1$Max--Planck--Institut f\"ur Festk\"orperforschung, D--70569 Stuttgart, Germany\\
$^2$Walther--Meissner--Institut f\"ur Tieftemperaturforschung, D--85748 Garching, Germany}\\
\end{center}
\section{A. Kinetic Theory for NCS.}
\renewcommand*{\theequation}{A.\arabic{equation}} 
\setcounter{equation}{0}
{{\bf \emph{Model description in equilibrium. }}
A non--centrosymmetric superconductor (NCS) is described in equilibrium by the Hamiltonian $\hat{\cal H}$, 
which is given by Eq.~(1) in the main text. 
In order to include the pairing correlations into the description, 
we extend Eq.~(1) to include the gap matrix $\vDelta_\vk$ as an off-diagonal element of 
an energy matrix $\underline{\xi}_\vk^0$ in Nambu space. In the presence of an antisymmetric spin--orbit coupling 
(ASOC), represented by the vector $\vgamma_\vk$, the $4\times4$ energy matrix has the following form 
in the spin representation:
\begin{equation}
\underline{\xi}_\vk^0~=~
\left(\begin{array}{cc}
\xi_\vk{\bf 1}+\tcbl{\vgamma_\vk}\cdot\vtau&\vDelta_{\vk} \\
\vDelta_{\vk}^\dagger & -[\xi_{-\vk}{\bf 1}+\tcbl{\vgamma_{-\vk}}\cdot\vtau]^T \\
\end{array}\right)  
\end{equation}
In order to account for the two--band structure occurring in NCS systems in the limit of large 
spin--orbit coupling, it is convenient to perform a unitary transformation of $\underline{\xi}_\vk^0$ 
into the {\it helicity--band basis} or simply {\it band basis}. The transformation from spin to 
band basis is described by the matrix $\vU_\vk$, which has the property
\begin{equation}
\vU^\dagger_\vk(\vgamma_\vk\cdot\vtau)\vU_\vk=\left\|\vgamma_\vk\right\|\vtau^3
\end{equation}
and which is obtained in the form of a SU(2) rotation
\begin{equation}
\vU_\vk={\rm e}^{-i\frac{\theta_\gamma}{2}\hat{\vn}_\gamma\cdot\vtau} \ ; \ 
\cos\theta_\gamma= \hat\vgamma_\vk\cdot\hat\vz \ ; \ 
\vn_\gamma=\frac{\vgamma_\vk\times\hat{\vz}}{\left\|\vgamma_\vk\times\hat{\vz}\right\|}
\end{equation}
that corresponds to a rotation in spin space into the $\hat{\vz}$--direction about the
polar angle $\theta_\gamma$ between $\vgamma_\vk$ and $\hat{\vz}$. Here, $\vtau$ denotes the 
vector of Pauli spin matrices. 
A straightforward extension of this transformation into Nambu space reads [S.1]
\begin{eqnarray}
\underline{\xi}_\vk^{(band)}\equiv\underline{U}_\vk^\dagger\underline{\xi}_\vk^0\underline{U}_\vk
=\left(\begin{array}{cccc}
\xi_{\vk+} & 0 & 0 & \Delta_{\vk+}\\
0 & \xi_{\vk-} & -\Delta_{\vk-} & 0 \\
0 & -\Delta_{\vk-}^* & -\xi_{\vk-} & 0 \\
\Delta_{\vk+}^* & 0 & 0 & -\xi_{\vk+}
\end{array}\right)
 \ ; \ \underline{U}_\vk=\left(\begin{array}{cc}
\vU_\vk & 0 \\
0 & \vU_{\vk}^* \\
\end{array}\right)
\end{eqnarray}
with the energy values $\xi_{\vk\mu}=\xi_{\vk}+\mu||\vgamma_\vk||$ and the gap functions 
$\Delta_{\vk\mu}=\Delta_s(T)+\mu\Delta_{tr}(T)f_\vk$ also given by Eq.~(3) in the main text. 
Introducing a band--index $\mu=\pm1$, one may write the equilibrium energy matrix in the 
band basis in the compact form:
\begin{eqnarray}
\underline{\xi}_{\vk\mu}^0=\left(\begin{array}{cc}
\xi_{\vk\mu} & \mu\Delta_{\vk\mu}\\
\mu\Delta_{\vk\mu}^* & -\xi_{-\vk\mu}
\end{array}\right)
\label{eq:EnergyComForm}
\end{eqnarray}
In analogy, one can find for the equilibrium density matrix:
\begin{eqnarray}
\underline{n}_{\vk\mu}^0=\left(\begin{array}{cc}
\frac{1}{2}-\xi_{\vk\mu}\theta_{\vk\mu} & -\mu\Delta_{\vk\mu}\theta_{\vk\mu}\\
-\mu\Delta_{\vk\mu}^*\theta_{\vk\mu} & \frac{1}{2}+\xi_{\vk\mu}\theta_{\vk\mu}
\end{array}\right)
\label{eq:DensityComForm}
\end{eqnarray}

{\bf\emph{Nonequilibrium Kinetic Equations. }}
The action of an external perturbation $\d\zeta=e\phi(\vq,\omega)+V_\vq\d n(\vq,\omega)$ 
leads to the deviation of the density matrix, as well as the energy matrix, from its equilibrium value. 
An NCS is now described in the band basis by 
a generalized momentum distribution function $\underline{n}_{\vp\vp^{\prime}}^\mu$ and an 
energy matrix $\underline{\xi}_{\vp\vp^{\prime}}^\mu$, respectively. A collisionless quantum 
dynamics is given by the von Neumann equation [see Eq.~(4) in the main text]:
\begin{equation}
  \hbar\omega\,\underline{n}_{\vp\vp^\prime}^\mu+\sum_{\vp^{\prime\prime}}
  [\underline{n}_{\vp\vp^{\prime\prime}}^\mu,\,\underline{\xi}_{\vp^{\prime\prime}\vp^{\prime}}^{\mu}]=0
\end{equation}
This equation can be linearized by using the ansatz of Eq.~(5) given in the main text. 
This leads to
\begin{equation}
\label{eq:MKEspinbase}
  \hbar\omega\delta\underline{n}_{\vk\mu}+\delta\underline{n}_{\vk\mu}
  \underline{\xi}_{\vk-\frac{\vq}{2}\mu}^{0} - 
  \underline{\xi}_{\vk+\frac{\vq}{2}\mu}^{0}\delta\underline{n}_{\vk\mu}=
  \delta\underline{\xi}_{\vk\mu}\underline{n}_{\vk-\frac{\vq}{2}\mu}^{0}
  -\underline{n}_{\vk+\frac{\vq}{2}\mu}^{0}\delta\underline{\xi}_{\vk\mu}
\end{equation}
with the equilibrium quasiparticle energy $\underline{\xi}_{\vk\mu}^0$ and 
the distribution function $\underline{n}_{\vk\mu}^0$  defined in 
Eqs.~\eqref{eq:EnergyComForm} and~\eqref{eq:DensityComForm},
respectively.
The momentum and frequency--dependent deviation from equilibrium can be defined in the appropriate way
as $2\times2$ matrices in the Nambu space:
\begin{equation}
\delta\underline{n}_{\vk\mu} =\left( \begin{array}{cc}
\delta n_{\vk\mu} & \mu\delta g_{\vk\mu}\\
\mu\delta g_{\vk\mu}^* & -\delta n_{-\vk\mu}
\end{array} \right)
\hspace{0.5cm}
{\rm and}
\hspace{0.5cm}
\delta \underline{\xi}_{\vk\mu} =
\left( \begin{array}{cc}
\delta\xi_{\vk\mu} & \mu\delta\Delta_{\vk\mu}\\
\mu\delta\Delta_{\vk\mu}^* & -\delta\xi_{-\vk\mu}
 \end{array} \right)
\end{equation}with $\d\xi_{\vk\mu}=\d\xi_{-\vk\mu}=\d\zeta$. Thus, the equation~\eqref{eq:MKEspinbase} 
represents a set of eight equations in the band basis [S.2]
(with the band index $\mu=\pm1$). Furthermore,
it is convenient to decompose the diagonal elements of the energy and density deviation matrices 
according to their parity with respect to $\vk\to-\vk$
\begin{equation}
 \begin{split}
  \d n_{\vk\mu}^{(s)}&=\frac{1}{2}\left(\d n_{\vk\mu}+s\d n_{-\vk\mu}\right)\\
  \d \xi_{\vk\mu}^{(s)}&=\frac{1}{2}\left(\d \xi_{\vk\mu}+s\d \xi_{-\vk\mu}\right)
 \end{split}
\end{equation}
 with the labeling $s=\pm1$. By analogy, the off--diagonal components are decomposed
 into their real and imaginary parts:
\begin{equation}
 \begin{split}
 \label{eq:Decomposition}
  \d g_{\vk\mu}^{(s)}&=\frac{1}{2}\left(\d g_{\vk\mu}\frac{\Delta_{\vk\mu}^*}{|\Delta_{\vk\mu}|}
  +s\frac{\Delta_{\vk\mu}}{|\Delta_{\vk\mu}|}\d g_{-\vk\mu}^*\right)\\
  \d \Delta_{\vk\mu}^{(s)}&=\frac{1}{2}\left(\d\Delta_{\vk\mu}\frac{\Delta_{\vk\mu}^*}{|\Delta_{\vk\mu}|}
  +s\frac{\Delta_{\vk\mu}}{|\Delta_{\vk\mu}|}\d\Delta_{-\vk\mu}^*\right)
 \end{split}
\end{equation}
where $\d\Delta^{(+)}_{\vk\mu}$ represents the amplitude fluctuations and $\Delta^{(-)}_{\vk\mu}$ 
the phase fluctuations of the order parameter. After these specifications 
the off--diagonal components of the Eq.~\eqref{eq:MKEspinbase}  simplify to [S.2]:
\begin{eqnarray} 
 \label{eq:OffDiagElPl}
  \d g_{\vk \mu}^{(+)}&=&-\left(\theta_{\vk\mu}+
 \frac{\omega^2-\left(\vv_{\vk\mu}\cdot\vq\right)^2-4\Delta_{\vk\mu}^2}{4\Delta_{\vk\mu}^2}
 \l_{\vk\mu}\right)\d\Delta_{\vk\mu}^{(+)} \\
 \delta g_{\vk\mu}^{(-)}+\theta_{\vk\mu}\delta\Delta_{\vk\mu}^{(-)}
&=&\frac{\omega\lambda_{\vk\mu}}{2\Delta_{\vk\mu}}\delta\zeta-
[\omega^2-(\vq\cdot\vv_{\vk\mu})^2]\lambda_{\vk\mu}\frac{\delta\Delta_{\vk\mu}^{(-)}}{4\Delta_{\vk\mu}^{2}}
\label{eq:OffDiagElMn}
\end{eqnarray}
whereas for the diagonal elements one gets:
\begin{eqnarray}
\label{eq:DiagElEqPl}
\delta n_{\vk\mu}^{(+)}&=&\left(\frac{\left(\vq\cdot\vv_{\vk\mu}\right)^2\varphi_{\vk\mu}}
{\omega^2-\left(\vq\cdot\vv_{\vk\mu}\right)^2}-\lambda_{\vk\mu}\right)\delta\zeta
+\omega\lambda_{\vk\mu}\frac{\delta\Delta_{\vk\mu}^{(-)}}{2\Delta_{\vk\mu}}\\
\delta n_{\vk\mu}^{(-)}&=&\frac{\omega\left(\vq\cdot\vv_{\vk\mu}\right)\varphi_{\vk\mu}}
{\omega^2-\left(\vq\cdot\vv_{\vk\mu}\right)^2}\delta\zeta
+\left(\vv_{\vk\mu}\cdot\vq\right)\lambda_{\vk\mu}\frac{\delta\Delta_{\vk\mu}^{(-)}}{2\Delta_{\vk\mu}} 
\label{eq:DiagElEqMn}
\end{eqnarray}
Equation~\eqref{eq:OffDiagElMn} describes the important relation between fluctuation of the 
pairing amplitude and the phase fluctuations of the superconducting order parameter 
[see Eq.~(6) in the main text]. The density response to a scalar perturbation $\d\zeta$ is given 
by Eq.~\eqref{eq:DiagElEqPl} [corresponding to Eq. (8) in the main text with 
$\d n_{\vk\mu}\equiv\d n_{\vk\mu}^{(+)}$]
\\

{\bf\emph{Conservation law. }}
One strength of the matrix kinetic equation approach lies in the straightforward physical interpretation of its results.
In addition, the gauge invariance of the whole theory can be demonstrated easily
if {\it all} phase fluctuation modes of the order parameter are properly taken into
account:
As one can see from Eqs.~\eqref{eq:DiagElEqPl}-\eqref{eq:DiagElEqMn} 
the density distribution functions $\delta n_{\vk\mu}^{(s)}$ are directly connected with 
the phase fluctuations of the order parameter $\delta\Delta_{\vk\mu}^{(-)}$. 
The combination of the results 
from Eqs.~\eqref{eq:DiagElEqPl}-~\eqref{eq:DiagElEqMn} yields together with the subsequent integration 
over the momentum space $\vk$ to the continuity equation
\begin{align}
\omega\delta n - \vq\cdot\vj =
\sum\limits_{\vp\mu} 
\lambda_{\vp\mu} \left\{ \left[ \omega^2 - \left( \vq\cdot\vv_{\vp\mu} \right)^2  \right]
\frac{\delta\Delta^{(-)}_{\vp\mu}}{2\Delta_{\vp\mu}} - \omega\delta\zeta \right\}
\end{align}
which at first glance displays a non--vanishing right-hand side. However, by using 
Eqs.~\eqref{eq:OffDiagElPl}-~\eqref{eq:OffDiagElMn} and the variation of the energy gap equation
\begin{equation}
  \delta\Delta_{\vk\mu}^{(-)}=\sum_{\vp\nu}\Gamma_{\vk\vp}^{\mu\nu}\delta g_{\vp\nu}^{(-)}
\end{equation}
one finds after a straightforward, but lengthy 
calculation:
\begin{equation}
 \omega\d n-\vq\cdot\vj=0\ .
\end{equation}
Thus, the particle conservation and, associated with it, the gauge invariance of the theory
are satisfied within the framework of the matrix kinetic theory.
\section{B. New collective modes}
\renewcommand*{\theequation}{B.\arabic{equation}} 
\setcounter{equation}{0}
The collective excitations of a non--centrosymmetric system can be obtained from the condition, that
the denominator of Eq. (10) vanishes, i.e. 
\begin{equation}
 \omega^4-
\left[\omega_{\rm P}^2(\hat\vq)+\omega_{\rm G}^2(\vq)
+\omega_{\rm L}^2(\vq)\right]\omega^2+\left[\omega_{\rm P}^2(\hat\vq)+\omega_{\rm G}^2(\vq)\right]
\omega_{\rm L}^{\prime 2}(\vq)=0
\label{eq:Quad}
\end{equation}
with $
\omega_{\rm G}^2(\vq)=\sum_{\mu}\lela\lambda_{\hat\vp\mu}(\vq\cdot\vv_{\vp\mu})^2\rira_{\rm FS}/\lambda
$
and
$
\omega_{\rm P}^2(\hat\vq)=4\pi n e^2\sum_\mu 3\lela\lambda_{\hat\vp\mu}(\hat\vq\cdot\hat\vp)^2\rira_{\rm FS}/m
$
. Here, we also use the abbreviation  
\begin{equation}
 \omega_{\rm L}^{\prime 2}(\vq)=\Lambda_0^2+\frac{\alpha_0\alpha_2\omega_{\vq 0}^2
 \omega_{\vq2}^2-\alpha_1^2\omega_{\vq 1}^4}{\left(\alpha_0\alpha_2-\alpha_1^2\right)\omega_{\rm G}^2(\vq)}
 \label{eq:LeggDisp}
\end{equation}
and
\begin{equation}
 \omega_{\rm L}^2(\vq)=\Lambda_0^2+\frac{\alpha_0\alpha_2
 \left(\omega_{\vq 0}^2+\omega_{\vq2}^2\right)-2\alpha_1^2\omega_{\vq 1}^2}
 {\left(\alpha_0\alpha_2-\alpha_1^2\right)} -\omega_{\rm G}^2(\vq)
 \label{eq:LeggDispUn}
\end{equation}
with the quantities $\alpha_n$, which are defined as
\begin{equation}
  \alpha_n=\sum_\mu \<\frac{\lambda_{\vk\mu}}{\Delta_{\vk\mu}^2}\left(\mu f_\vk\right)^n\>_{\rm FS}
  \label{eq:DefAlpha}
\end{equation}
together with:
\begin{equation}
\omega_{\vq n}^2=\frac{1}{\alpha_n}
\sum_\mu \<\frac{\lambda_{\vk\mu}}{\Delta_{\vk\mu}^2}\left(\vq\cdot\vv_{\vk\mu}\right)^2
\left(\mu f_\vk\right)^n\>_{\rm FS} 
\label{eq:DefOmega}
\end{equation}
and the Leggett mass $\Lambda_0^2\equiv\omega_{\rm L}^2(\vq=0)$.
Thus, Eq.~\eqref{eq:Quad} is a quadratic equation with respect to $\omega^2$ 
with the solutions
\begin{equation}
 \omega^2_{1,2}=\frac{1}{2}\left[\omega_{\rm P}^2(\hat\vq)+\omega_{\rm G}^2(\vq)
+\omega_{\rm L}^2(\vq)\pm\left(\omega_{\rm P}^2(\hat\vq)+\omega_{\rm G}^2(\vq)
+\omega_{\rm L}^2(\vq)\right)\sqrt{1-4\frac{\left[\omega_{\rm P}^2(\hat\vq)+\omega_{\rm G}^2(\vq)\right]
\omega_{\rm L}^{\prime 2}(\vq)}
{\left(\omega_{\rm P}^2(\hat\vq)+\omega_{\rm G}^2(\vq)
+\omega_{\rm L}^2(\vq)\right)^2}}\ \right].
\end{equation}
This result can be further simplified by using a Taylor expansion of the square root. Therefore, 
by considering terms up to second order in $\left|\vq\right|$ one gets:
\begin{equation}
  \omega_{1,2}^2=\omega_{\rm P}^2(\hat\vq)+\omega_{\rm G}^2(\vq)
+\omega_{\rm L}^2(\vq)\pm\frac{\left[\omega_{\rm P}^2(\hat\vq)+\omega_{\rm G}^2(\vq)
\right]\omega_{\rm L}^{\prime 2}(\vq)}{\omega_{\rm P}^2(\hat\vq)+\omega_{\rm G}^2(\vq)
+\omega_{\rm L}^2(\vq)}
\label{eq:Res}
\end{equation}
In the absence of the long--range Coulomb interaction (i.e. for the case $\omega_{\rm P}^2(\hat\vq)=0$) 
one finds from Eq.~\eqref{eq:Res} following result for the collective modes [see Eq.~(11)
in the main text]:
\begin{equation}
\begin{array}{cccl}
  \omega_1^2&=&\omega_{\rm G}^2(\vq)+{\cal O}\left(\dfrac{\omega_{\rm G}^4(\vq)}
  {\omega_{\rm L}^2(\vq)}\right)&\quad \text{Gauge mode}\\
  \omega_2^2&=&\omega_{\rm L}^2(\vq)+{\cal O}\left(\dfrac{\omega_{\rm G}^4(\vq)}
  {\omega_{\rm L}^2(\vq)}\right)&\quad \text{Leggett mode}
\end{array}
\end{equation}
The Coulomb interaction leads to the renormalization of this result [see Eq.~(12)
in the main text]:
\begin{equation}
\begin{array}{ccll}
  \omega_1^2&=&\omega_{\rm P}^2(\hat\vq)+\omega_{\rm G}^2(\vq)
+\omega_{\rm L}^2(\vq)-\omega_{\rm L}^{\prime 2}(\vq)+{\cal O}\left(\dfrac{\omega_{\rm L}^2(\vq)}
  {\omega_{\rm P}^2(\hat\vq)}\right)&\quad \text{Plasma mode}\\
  \omega_2^2&=&\omega_{\rm L}^{\prime 2}(\vq)+{\cal O}\left(\dfrac{\omega_{\rm L}^2(\vq)}
  {\omega_{\rm P}^2(\hat\vq)}\right)&\quad \text{Leggett mode}
\end{array}
\end{equation}
Thus, the mass of the Leggett mode remains unaffected by this process, but its dispersion 
is changed. In the limiting case of small $\vq$ the dispersion modification is, however, negligible.
\section{C. Anderson--Higgs mechanism}
\renewcommand*{\theequation}{C.\arabic{equation}} 
\setcounter{equation}{0}
\renewcommand*{\thefigure}{C.\arabic{figure}} 
\setcounter{figure}{0}
In order to discuss the Anderson--Higgs mechanism for the Leggett mode in non--centrosymmetric 
superconductors we consider the difference between the Coulomb--renormalized 
Leggett mode $\omega_{\rm L}^\prime$ defined in Eq.~\eqref{eq:LeggDisp} and 
its unrenormalized counterpart $\omega_{\rm L}$ defined in Eq.~\eqref{eq:LeggDispUn}:
\begin{equation}
 \omega_{\rm L}^2(\vq)-\omega_{\rm L}^{\prime 2}(\vq)=\frac{\alpha_0\alpha_2
 \left(\omega_{\vq 0}^2+\omega_{\vq2}^2\right)-2\alpha_1^2\omega_{\vq 1}^2}
 {\left(\alpha_0\alpha_2-\alpha_1^2\right)} -\omega_{\rm G}^2(\vq)
-\frac{\alpha_0\alpha_2\omega_{\vq 0}^2
 \omega_{\vq2}^2-\alpha_1^2\omega_{\vq 1}^4}{\left(\alpha_0\alpha_2-\alpha_1^2\right)\omega_{\rm G}^2(\vq)}
 \label{eq:DispDif}
\end{equation}
For simplicity we make following assumptions: 
(i) low temperature limit ($T\to0$); (ii) isotropic spin--orbit coupling $f_\vk=1$ [corresponding to 
the leading order of 
$\gamma_\vk$ for the cubic point group $O(432)$]. A generalization beyond these approximations is, 
however, straightforward. With these assumptions, 
equation~\eqref{eq:DispDif} simplifies to
\begin{equation}
 \omega_{\rm L}^2(\vq)-\omega_{\rm L}^{\prime 2}(\vq)=\frac{2\alpha_0^2
 \omega_{\vq 0}^2-2\alpha_1^2\omega_{\vq 1}^2}
 {\left(\alpha_0^2-\alpha_1^2\right)} -\omega_{\rm G}^2(\vq)
-\frac{\alpha_0^2\omega_{\vq 0}^4
 -\alpha_1^2\omega_{\vq 1}^4}{\left(\alpha_0^2-\alpha_1^2\right)\omega_{\rm G}^2(\vq)}
 \label{eq:DispDifSimpl}
\end{equation}
with $\omega_{\rm G}^2=\frac{1}{6}\sum_{\mu}v_{{\rm F}\mu}^2|\vq|^2$ and $v_{{\rm F}\mu}$ being
the Fermi velocity on the band $\mu=\pm1$. 
By using the definitions~\eqref{eq:DefAlpha} and \eqref{eq:DefOmega} 
the equation~\eqref{eq:DispDifSimpl} can be further simplified. Thus, after straightforward calculations 
one obtains:
\begin{equation}
 \omega_{\rm L}^2(\vq)-\omega_{\rm L}^{\prime 2}(\vq)=\omega_{\rm G}^2(\vq)
-\frac{\<\left(\vq\cdot\vv_{\vk+}\right)^2\>_{\rm FS}\<\left(\vq\cdot\vv_{\vk-}\right)^2\>_{\rm FS}}
{\omega_{\rm G}^2(\vq)}
\end{equation}
with the Fermi surface average $\<\dots\>$ defined as 
$\<z(\vk^\prime)\>_{\rm FS}=\int d\phi\int d\theta z(\vk^\prime) \sin\theta$ for 
a given function $z(\vk^\prime)$.
Finally, assuming the same DoS on both bands, i.e. $N_\mu=N_0$, and almost similar Fermi 
velocities, i.e. $(v_{{\rm F}+}-v_{{\rm F}-})\ll v_{\rm F}$
with $v_{\rm F}=\max_{\mu}v_{{\rm F}\mu}$, one obtains after performing the integration:
%
%
\begin{equation}
 \frac{\omega_{\rm L}^2(\vq)-\omega_{\rm L}^{\prime 2}(\vq)}{v_{\rm F}^2|\vq|^2}\approx\frac{1}{6}\frac{
 \left(v_{{\rm F}+}^2-v_{{\rm F}-}^2\right)}{v_{{\rm F}+}^2+v_{{\rm F}-}^2}
 \frac{\left(v_{{\rm F}+}^2-v_{{\rm F}-}^2\right)}{v_{\rm F}^2}\ll 1
\end{equation}
%
%
%
%
%
%
%
%
\begin{center}
\rule{4cm}{0.9pt}
\end{center}
[S.1] A. Vorontsov, I. Vekhter, and M. Eschrig. {\it Physica B} {\bf 403}, 1095 (2008)\newline
[S.2] E. Bauer, M. Sigrist, eds., {\it Non--centrosymmetric superconductors} (Springer, Heidelberg, 2012)
}
%


%
%
\end{document}